\documentclass[prd,twocolumn,showpacs,amsmath,amssymb,superscriptaddress,preprintnumbers,nofootinbib]{revtex4}
\usepackage{amssymb}
\usepackage{amsmath}
\usepackage{graphicx}
\usepackage[normalem]{ulem}
\usepackage{cancel}

\normalbaselines
\begin{document}

\title{Spin-axion coupling}

\author{Alexander B. Balakin}
\email{Alexander.Balakin@kpfu.ru} \affiliation{Department of
General Relativity and Gravitation, Institute of Physics, Kazan
Federal University, Kremlevskaya str. 18, Kazan 420008, Russia}

\author{Vladimir A. Popov}
\email{Vladimir.Popov@kpfu.ru} \affiliation{Department of
General Relativity and Gravitation, Institute of Physics, Kazan
Federal University, Kremlevskaya str. 18, Kazan 420008, Russia}

\preprint{\bf version 1.2}

\begin{abstract}
We establish a new covariant phenomenological model, which describes an influence of pseudoscalar (axion) field on spins of test massive particles. 
The model includes general relativistic equations of particle motion and spin evolution
in background pseudoscalar (axion), electromagnetic and gravitational fields. It describes both
the direct spin-axion coupling of the gradient type and
indirect spin-axion interaction mediated by electromagnetic fields.
Special attention is paid to the direct spin-axion coupling caused by the gradient of the pseudoscalar (axion) field. We show that it describes a spin precession, when the pseudoscalar (axion) field is inhomogeneous and/or non-stationary.
Applications of the model, which correspond to the three types of four-vectors attributed to the gradient of the pseudoscalar (axion) field
(time-like, space-like, and null), are considered in detail.
These are the spin precessions induced by relic cosmological axions, axions distributed around spherically symmetric static objects, and  axions in a gravitational wave field, respectively.
We discuss features of the obtained exact solutions and some general properties of the axionically induced spin rotation.
\end{abstract}


\maketitle

\section{Introduction}\label{sec - intro}

The terms \emph{angular moment} and \emph{spin} are considered in
theories of gravity in two main aspects. First, we face with them,
when we study the gravitational field formed by a single rotating
extended body (e.g., black holes, neutron stars, etc. \cite{B1,B2}),
or a system of bodies rotating around an attracting center (e.g.,
stars in spiral galaxies \cite{B3,B4}). Second, we deal with the
polarization or spin, when we investigate the gravitationally
induced dynamics of test particles, which possess vector (or
pseudo-vector, respectively) degrees of freedom. This sector of
investigations is usually connected with experiments in
gravitational physics (see, e.g., \cite{B5,B6} and references therein).

According to the
standard terminology the rotating extended body or point-like
particle can be described by an anti-symmetric tensor of total
moment $S_{ik}$, which can be decomposed as
$S_{ik}{=}\delta^{mn}_{ik}L_mU_n{-}\epsilon_{ikmn}S^mU^n $, using
the unit velocity four-vector $U^n$ of the body. The quantity
$L^m$ can be indicated as a four-vector of the orbital moment, and
the (pseudo)vector $S^m$ plays the role of a classical spin
four-vector. The term $\delta^{mn}_{ik}$ is the four-indices
Kronecker tensor and $\epsilon_{ikmn}$ is the Levi-Civita
(pseudo)tensor. Evolutionary equations for the total moment
$S_{ik}$ and/or for its constituents $L^m$ and $S^m$ have been
investigated in various contexts by many authors.
The list of obtained results is very long, and we would like to
attract the attention to the following two ones only.

In 1959 Bargmann, Michel and Telegdi obtained the covariant equations for
classical spin particles with anomalous magnetic
moment~\cite{BMT}. This model (BMT-model, for short) describes a
spin evolution in the framework of Special Relativity as a
generalization of the non-relativistic theory elaborated by
Thomas~\cite{Thomas}, Frenkel~\cite{Frenkel} and Bloch~\cite{Bloch}. Being completed by
the spin-curvature coupling terms, introduced earlier by Mathisson~\cite{Mathisson} and
Papapetrou~\cite{Papapetrou}, this model became a starting point for
numerous investigations of spin particle dynamics in General
Relativity (see, e.g.,~\cite{B7,B8,B9,B10,B11,B12,B13,B14} and references therein).
When it was necessary to include electromagnetic interactions, these investigations used the Faraday-Maxwell
electrodynamics. Scientific events of two last decades attracted
the attention to its generalization called \emph{Axion
Electrodynamics}. In the light of the hypothesis about axionic
nature of the dark matter this theory seems to be more appropriate
for describing the cosmic electrodynamics than the standard
Faraday-Maxwell theory.

The story, how axions were associated with dark matter particles, is well-known.
The dark matter, a cosmic substance, which neither emits nor
scatters the electromagnetic radiation, is assumed to accumulate
about 23\% of the Universe energy. The mass density distribution
of the dark matter is presented in the astrophysical catalogues as
a result of observations and theoretical simulations (see, e.g.,
\cite{DM1,DM2,DM3,DM4,DM5} for details, review and references). The origin
of the dark matter is not yet established. One of the most
attractive hypothesis links the dark matter with massive
pseudo-Goldstone bosons. The particles of this type were
postulated in 1977 by Peccei and Quinn \cite{PQ} in order to solve
the problem of strong $CP$-invariance, and were introduced into
the high-energy physics as new light bosons by Weinberg
\cite{Weinberg} and Wilczek \cite{Wilczek} in 1978. Later these
pseudo-bosons were indicated as {\it axions} and now they are considered as
the most appropriate candidate for dark matter particles \cite{A1,A2,A3,A4,A5,A6,A7,A8,A9}.

The description of these axions in terms of the field theory is
based on the introduction of a pseudoscalar field $\phi$, which is
assumed to interact with electromagnetic and $SU(2)$ - symmetric
gauge fields. The theory of interaction between electromagnetic
and pseudoscalar fields was elaborated by Ni in 1977 \cite{WTN77}.
Due to the works of Sikivie (see, e.g., \cite{Sikivie}) we recognize
now this sector of science as \emph{axion electrodynamics}; a
number of effects such as the axionically induced optical activity
\cite{OA1,OA2}, magneto-electric cross-effects \cite{ME1,ME2}, etc., were
predicted in the framework of this theory.

Thus, on the one hand, the axions are associated in our mind with
the cosmic dark matter, which regulates key gravitational
processes in our Universe, for instance, provides specific
(flat) distribution of star velocities in the spiral galaxies (see, e.g.,
 \cite{NFW}). On the other hand, the axions produce the photon polarization rotation,
in other words, the axionic dark matter forms an electrodynamically
active chiral cosmic medium. Clearly, we could try to find
fingerprints of the axionic dark matter in the electromagnetic
signals associated with astrophysical and geophysical phenomena
\cite{K1,K2,K3,K4,K5}.

Returning to the problem of influence of the (axionic) dark matter
on the orbital moment, polarization and spin, we can say the
following. First, definitely, the dark matter influences the
orbital moment of star gas in spiral galaxies, providing the
observed flat profiles of the velocity distribution \cite{NFW}.
Second, the axionic dark matter is
predicted to rotate the photon polarization \cite{OA1}. Third,
due to the first and second arguments, one can expect that the axionic
dark matter should also influence the particle spin.

In fact, the theoretical discussions about a new branch in the axion physics, i.e., the theory of \emph{spin-axion coupling}, were opened in 80s of the last century; it was a discussion concerning the axionically induced spin dependent long range forces \cite{WLCK1984}. The long range spin dependent forces were in a focus of experimental investigations (see, e.g., \cite{Ramsey}), and this circumstance gave the impetus for development of new experimental programs devoted to detection of axions, based on their interactions with nuclear spins, electric and magnetic moments, etc. (see, e.g., \cite{E1,E2,E3,E4,E5,E6,E7,E8,E9} and references therein).
This branch of axion physics was indicated as spin-axion coupling by analogy with the axion-photon and the axion-gluon couplings. Nevertheless the term spin-axion coupling is rather wide and two theoretical approaches have to be distinguished. Experimental works in the terrestrial laboratories deal with bonded spin particles in the specific material media, and these \emph{bonded} particles are, in average, at rest with respect to these media. For these purposes the non-relativistic   formalism of condensed matter physics is adequate and describes correctly the models of spin dynamics.
When one deals with \emph{free} spin particles, which move with high velocity (e.g., in storage rings) or in a vicinity of sources of strong gravitational and axion fields (e.g., near axion stars), the covariant formalism of high energy physics is necessary for description of the model.

From the mathematical point of view, our goal is to develop the covariant version of the theory of spin-axion coupling, and we consider extended dynamic equations for relativistic spin-particle. The corresponding extensions include axionic modifications of the Bagrmann-Michel-Telegdy force-like term, and gradient-type modifications of a rotatory term.
From the physical point of view, adding the third entity, the axion, to the interacting
pair of photon and charged spin particle, one can expect that three channels
of interaction will be activated: first, the already known direct
axion-photon coupling; second, the direct spin-axion coupling;
third, the indirect spin-axion coupling mediated by the
axion-photon interaction. In this paper we intend to discuss in detail two last channels of
the photon-axion-spin interactions.

The paper is organized as follows. In Section~\ref{sec - photon polarization} we analyze some
details of the well-known model of axion-photon coupling to
motivate our new model of spin-axion coupling. In particular, we
derive an equation of the photon polarization precession induced
by a gradient of the pseudoscalar (axion) field. In Section~\ref{sec - particle dynamics} we
generalize the results of Section~\ref{sec - photon polarization} for massive particles and
reconstruct phenomenologically equations of spin-particle dynamics
in the axion environment. We divide the obtained terms into three
types: first, the terms indicated as axionic generalizations of
the Bargmann-Michel-Telegdy terms; second, the terms describing
direct spin-axion coupling; third, the nonminimal terms. In
Section~\ref{sec - applications} we consider four applications of the model with the
direct spin-axion coupling. In Subsection~\ref{sec - relic DM} we obtain an exact
solution to the dynamic equations for the massive spin
particle, which moves straightforwardly and interacts with the relic cosmic axions. In Subsection~\ref{sec - StorageRing} we apply the formalism to the model of relativistic charged spin particle motion in a storage ring with magnetic field. In Subsection~\ref{sec - axion star}
we discuss the spin-axion coupling in the static spherically
symmetric gravitational background. We studied in detail two
cases, describing radial and circular particle motion,
respectively. The exact solutions for the spin precession in the
field of pp-wave gravitational field provided by axions are
presented in Subsection~\ref{sec - GW}. In Section~\ref{sec - discussion} we discuss the common
details of the axionically induced spin precession, which were
revealed in all three submodels.

\section{Classical analogy: photon polarization rotation induced
by the axion field}\label{sec - photon polarization}

In 1977 Ni \cite{WTN77} obtained equations for the
electromagnetic field coupled to a pseudoscalar field $\phi$. The
initial form of the currentless electrodynamic equations was the
following:
\begin{equation}
\nabla_l \left[F^{kl} + \phi F^{*kl} \right] = 0\,.  \label{int2}
\end{equation}
This equation is derived by variation with respect to the potential four-vector $A_i$
of the electromagnetic part $S_{({\rm EM})}$ of
the total action functional, where
\begin{equation}
S_{({\rm EM})} = \int d^4x \sqrt{-g} \left[\frac{1}{4}
F_{mn}F^{mn} + \frac{1}{4} \phi  F^{*}_{mn}F^{mn} \right].
\label{int1}
\end{equation}
Here $F_{mn}$ is the Maxwell tensor, and $F^{*kl}=\frac12
\epsilon^{klmn}F_{mn}$ is its dual tensor. As usual,
$\epsilon^{klmn} \equiv \frac{E^{klmn}}{\sqrt{-g}}$ is the
Levi-Civita tensor, based on the completely anti-symmetric
Levi-Civita symbol $E^{klmn}$ with $E^{0123}{=}1$. The Maxwell
tensor $F_{mn}$ is connected with the electromagnetic potential
$A_k$ in the regular way:
\begin{equation}
F_{mn}= \nabla_m A_n - \nabla_n A_m \,, \label{int4}
\end{equation}
so, that the well-known relationship
\begin{equation}
\nabla_l F^{*kl}=0  \label{int3}
\end{equation}
converts into the identity. Due to (\ref{int3}) the basic equation
(\ref{int2}) simplifies as follows:
\begin{equation}
\nabla_l F^{kl} = - F^{*kl} \nabla_l \phi \,. \label{int5}
\end{equation}
In order to transform Eq.~(\ref{int5}) to an one-photon form we
use the standard procedure. First, we follow the approximation of
geometrical optics, and represent the potential $A_m$ as
\begin{equation}
A_m = a_m e^{i \Psi} \,. \label{int6}
\end{equation}
Here $\Psi$ is a rapidly varying phase, the so-called eikonal. Its
gradient four-vector $k_j \equiv \nabla_j \Psi$ has a sense of a
wavelength four-vector of the photon. The amplitude $a_m$ and its
derivatives are assumed to be slowly varying functions, i.e.,
${\rm max}\left|\nabla_j a_m \right| \ll {\rm min}\left|k_j a_m
\right|$. In these terms the Lorentz condition
\begin{equation}
\nabla_m A^m = 0  \label{g4}
\end{equation}
gives two relations. The first one,  $a_m k^m =0$, appears in the
first order approximation and means that the polarization is
orthogonal to the wavelength four-vector. In the next order we
obtain $\nabla_m a^m =0$. The leading order approximation in
Eq.~(\ref{int5}) in the context of (\ref{int6}) yields $k_m k^m
=0$, the ordinary eikonal equation, implying that the photons
travel along null geodesics. The tangent  vector to the photon
world-line can be defined as $k^m = \hbar \frac{\ dx^m}{d\tau}$,
where $\tau$ is an affine parameter along the line, and $\hbar$ is
the Planck constant.

The differential consequence of the eikonal
equation $k_m k^m =0$ is $k^m \nabla_n k_m =0$. Since the quantity
$k^m$ is the gradient four-vector, the relation $\nabla_m
k_{n}=\nabla_n k_{m}$ is valid, so we obtain additionally that
\begin{equation}
k^m \nabla_m k^n =0\,. \label{addg1}
\end{equation}
For an illustration of main idea we assume that the wavelength
vector $k^m$ is divergence-free, i.e., $\nabla_m k^m =0$ (or
equivalently, $\nabla^m \nabla_m \Psi =0$). This can be realized
in many models, e.g, in the Minkowski space-time, in the
space-time with plane-wave symmetry. Then the first order
approximation in Eq.~(\ref{int5}) gives the following equation for the
evolution of the amplitude $a^j$:
\begin{equation}
k^m \nabla_m a^j = \frac12 \epsilon^{jlpq}\nabla_l \phi \  a_p k_q
\,. \label{g5}
\end{equation}
The convolution of this equation with $a_j$ yields
\begin{equation}
\frac12 k^m \nabla_m (a^ja_j) = 0 \,, \label{g7}
\end{equation}
i.e., $a^ja_j \equiv -a^2 $ is constant along the photon
world-line with the tangent  vector $k^l$. This fact allows us to
introduce the unit  space-like polarization four-vector $\xi^j$ as
follows:
\begin{equation}
a^j = a \ \xi^j \,, \quad  \xi^j \xi_j = - 1 \,,\label{g8}
\end{equation}
and to exclude the amplitude $a$ from Eq.~(\ref{g5}):
\begin{equation}
k^m \nabla_m \xi^j = \frac12 \epsilon^{jlpq}\nabla_l \phi  \ \xi_p
k_q \,. \label{g9}
\end{equation}
Using the covariant differential operator
\begin{equation}
\frac{D}{D\tau} \equiv \frac{\ dx^m}{d\tau} \nabla_m \,,
\label{g101}
\end{equation}
one can rewrite Eqs.~(\ref{addg1}) and (\ref{g9}) as a pair of
basic equations of photon evolution: first, the dynamic equation
\begin{equation}
\frac{D k^j}{D\tau} = 0 \,, \label{g12}
\end{equation}
and second, the equation of polarization rotation
\begin{equation}
\frac{D\xi^j}{D\tau} = \frac12 \epsilon^{jl}_{\ \ pq}\nabla_l \phi
\ \xi^p \  \frac{dx^q}{d\tau} \,. \label{g10}
\end{equation}
It is clear, that in terms of one-photon description three scalar
quantities,  $k^m k_m$, $k_m \xi^m$, and  $\xi^m \xi_m$, are the
integrals of motion for the set of equations (\ref{g12}),
(\ref{g10}), i.e., they remain constant along the photon
world-line, $k^m k_m=$const$=0$, $k_m \xi^m=$const$=0$, $\xi^m \xi_m
=-1$.

There is an obvious analogy between equations of motion for the
massless  photons and massive fermions. Based on this analogy, we
can replace the polarization four-vector  $\xi^j$ by the spin
four-(pseudo)vector $S^j$, introduce the time-like particle
momentum four-vector $p^k = m U^k=m \frac{\ dx^k}{ds}$ instead
of the null wave four-vector $k^j$ (here and below we consider
the system of units, in which $c=1$). As a natural extension of
equations (\ref{g12}) and (\ref{g10}) we obtain the following
evolutionary equations for the fermion particle:
\begin{equation}
\frac{D p^j}{D\tau} = 0 \,, \qquad  \frac{DS^j}{D\tau} =
\frac{\alpha}{2m} \epsilon^{jl}_{\ \ pq}\nabla_l \phi \ S^p \ p^q
\,, \label{g107}
\end{equation}
where $\alpha$ is some parameter introduced phenomenologically.
The presented analogy could explain the appearance of the
rotational term in the right-hand side of the equation
(\ref{g107}). However, some fermions possess electric charges and thus
interact with electromagnetic fields, and a phenomenological
generalization of equations (\ref{g12}) and (\ref{g10}) requires
more sophisticated efforts. We will take into account these
interactions in the next section.

\section{Spin-particle dynamics in an axion environment}
\label{sec - particle dynamics}

\subsection{Basic equations}\label{sec - basic eqs}

Let us consider the evolution of relativistic point particle with
an electric charge and a spin four-vector. Dynamic equations
for the particle momentum $p^{i}$ and for the spin four-vector
$S^{i}$ can be written as:
\begin{equation}
\frac{D p^i}{D\tau} = {\cal F}^i \ , \quad \frac{D S^i}{D\tau} =
{\cal G}^i \,, \label{1}
\end{equation}
i.e., the rates of change of the given quantities are
predetermined by the corresponding force-like terms, ${\cal F}^i $
and ${\cal G}^i $, respectively. Let us present, first, three
general properties of the four-vectors ${\cal F}^i $ and ${\cal
G}^i $.

\noindent {\it (i)}. The mass of the particle, $m$, defined from
the normalization law $p^i p_i = m^2c^2$,  is assumed to be
conserved quantity, providing the four-vector ${\cal F}^i$ to be
orthogonal to the momentum:
\begin{equation}
 p_i {\cal F}^i  =  p_i \frac{D p^i}{D\tau} = \frac12\frac{D}{D\tau} (p_i p^i) =0
 \,.
\label{s1}
\end{equation}

\noindent {\it (ii)}. Similarly, we assume that the scalar square
of the space-like spin four-vector is constant, i.e.,  $S^i S_i =$const$= - {\cal S}^2$. Then using the second equation from (\ref{1}) we
obtain that
\begin{equation}
 S_i {\cal G}^i  =  S_i \frac{D S^i}{D\tau} = \frac12\frac{D}{D\tau} (S_i S^i)
 = - \frac12\frac{D}{D\tau} {\cal S}^2 =0
 \,,
\label{s2}
\end{equation}
or in other words, that the force-like term ${\cal G}^i$ is
orthogonal to the spin four-vector.

\noindent {\it (iii)}. Finally, we assume that the spin
four-vector is orthogonal to the momentum four-vector, $S^ip_i =$const$=0$; then one obtains that
\begin{equation}
\frac{D}{D\tau} (p^i S_i) = 0 \ \ \Rightarrow \ {\cal F}^i  S_i + {\cal
G}_i p^i =0 \,. \label{111}
\end{equation}
In the minimal theory the quantities ${\cal F}^i $ and ${\cal G}^i
$ depend on the particle momentum $p^k$, spin $S^l$, Maxwell
tensor $F_{mn}$ and its dual $F^{*}_{mn}$, as well as, on the
pseudoscalar (axion) field $\phi$ and its gradient four-vector
$\nabla_k \phi$. In the nonminimally extended theory the
quantities ${\cal F}^i $ and ${\cal G}^i $ can include the Riemann
tensor $R^{i}_{\ klm}$, Ricci tensor $R_{ik}$ and Ricci scalar
$R$; also the covariant derivatives of the Maxwell tensor,
$\nabla_s F_{mn}$, and of the Riemann tensor $\nabla_s R_{ikmn}$
can appear in the decomposition of ${\cal F}^i $ and ${\cal G}^i$.
Three equations (\ref{s1}), (\ref{111}) and (\ref{s2}) are
satisfied identically, when
\begin{equation}
{\cal F}^i  = \omega^{ik} p_k  \,, \quad {\cal G}^i = \omega^{ik} S_k \,,
\label{BMT1}
\end{equation}
where $\omega^{ik}$ is an arbitrary anti-symmetric tensor,
$\omega^{ik}= - \omega^{ki}$. As a first step, we remind to the
reader, what is this tensor for the most known example, the
Bargmann-Michel-Telegdi model \cite{BMT}.

\subsection{Bargmann-Michel-Telegdi model}\label{sec - BMT}

Let us consider the Bargmann-Michel-Telegdi model,  for which the
relevant force-like term ${\cal F}^i $ is the Lorentz force
\begin{equation}
{\cal F}^i = \frac{e}{m}F^i_{ \ k} \ p^{k} \,, \label{2}
\end{equation}
and ${\cal G}^i$ is of the form
\begin{equation}
{\cal G}^i = \frac{e}{2 m} \left[ \ g F^i_{\ k} \ S^{k} +
\frac{(g - 2)}{ m^2}  p^i F_{kl} S^k p^l \right]. \label{5}
\end{equation}
Here $g$ denotes the so-called  $g$-factor; the case $g\neq 2$
means that the particle possesses an anomalous magnetic moment.
Clearly, these terms can be written in the form (\ref{BMT1}), when
\begin{equation}
\omega^{ik} = \omega^{ik}_{(0)} {=} \frac{e}{2m} \left[g F^{ik} +
\frac{\left(g{-}2 \right)}{m^2} \delta^{ik}_{mn} p_{j}F^{jm}p^n
\right] , \label{BMT2}
\end{equation}
where
$\delta^{ik}_{mn}=\delta^{i}_{m}\delta^{k}_{n}-\delta^{i}_{n}\delta^{k}_{m}$
is the four-indices Kronecker tensor. The important detail of this
model is that the force ${\cal F}^i $ does not contain the spin
four-vector, and the term ${\cal G}^i$ is linear in $S^k$. There
is a simple motivation of such model construction. In the
quasi-classical approach one uses the decomposition of microscopic
equations with respect to small quantity $\hbar$, the Planck
constant. Although the spin of particle enters  the microscopic
equations in the product $\hbar \cdot S$, the Planck constant is not involved into the
classical dynamic equations. Therefore one has to exclude the quantity $\hbar \cdot S$ from (\ref{BMT2}) to provide that both left-hand and
right-hand sides of Eqs.~(\ref{1}) are of the same order in $\hbar$,
and the multiplier $\hbar$ can be eliminated.

\subsection{Electrically charged spin particle in electromagnetic and axion fields}
\label{sec - charge in EM and Ax fields}

\subsubsection{Reconstruction of the force-type sources}\label{sec - reconstruction}

Keeping in mind the general relationships (\ref{BMT1}) we
reconstruct the tensor $\omega^{ik}$ using the following ansatz:

\noindent
A) The tensor   $\omega^{ik}$ does not contain the four-vector $S^i$.

\noindent
B) The tensor $\omega^{ik}$ is up to the first order in the
Maxwell tensor $F_{ik}$.

\noindent C) The tensor $\omega^{ik}$ is linear in the
pseudoscalar (axion) field $\phi$, or in its gradient four-vector
$\nabla_i \phi$.

\noindent D)
The tensor $\omega^{ik}$ is linear in the Riemann
tensor.

The first two requirements are the same as in the
Bargmann-Michel-Telegdi model. The third point is a new detail,
which appears as a natural extension of this model for  the case
of particles interacting with the pseudoscalar field. The last
point implies that the model can be minimal, when $R^i_{\ klm}$
does not enter the tensor $\omega^{ik}$, and nonminimal, when
there are terms containing the Riemann tensor and its linear
convolutions. Below we consider all the appropriate constructions,
which satisfy the requirements A)-D) and can be added to the
tensor $\omega^{ik}_{(0)}$ (\ref{BMT2}).

Minimal (curvature free) terms linear in the axion field  $\phi$
itself can be represented as follows:
\begin{equation}
\omega^{ik}_{(1)} {=} \frac{e \lambda}{2m} \phi \ \left[g_{A}
F^{*ik} {+} \frac{\left(g_{A}{-}2 \right)}{m^2}
\delta^{ik}_{mn} p_{j}F^{*jm}p^n \right]. \label{d2}
\end{equation}
In fact, $\omega^{ik}_{(1)}$ can be obtained from
$\omega^{ik}_{(0)}$ with  replacements $F_{mn} \to F^{*}_{mn}$ and
$g \to g_{A}$, where the coupling constant $g_{A}$ is an axionic
analog of $g$-factor. Dimensionless parameter $\lambda$ is equal
to one, if the Nature admits this coupling term, and $\lambda=0$,
if it does not admit.

Minimal terms linear in the gradient four-vector of the axion
field can be written as follows:
\begin{eqnarray}
\omega^{ik}_{(2)} &=&
\frac{e \mu}{2m} p^l \nabla_l \phi \biggl[g_{G} F^{*ik}  + \frac{(g_{G}-2)}{m^2}\delta^{ik}_{mn} p_{j}F^{*jm}p^n +
\nonumber
\\ &&
+ \, \omega_{23} \epsilon^{ikjn} p_n   p^{s}F_{js} \biggr] {+}
\label{d3}
\\ &&
+ \frac{\nabla_l \phi}{m}\left[ \omega_{24} \epsilon^{ikmn} p_n
F_{m}^{\ \ l}  {+} \omega_{25} \epsilon^{ikln} p_n \right].
\nonumber
\end{eqnarray}
Here, in addition to the axionic analog of the $g$-factor $g_{A}$,
we introduced its gradient-type analog $g_{G}$. The constant
$\mu$ in (\ref{d3}) plays the same role as the constant $\lambda$
in (\ref{d2}). Other constants have no direct analogs, and we
indicated them as $\omega_{23}$, $\omega_{24}$, etc., where the first
index 2 is an indicator that the decomposition relates to the
term $\omega^{ik}_{(2)}$. The last term in (\ref{d3}) is a unique
element of the presented irreducible decomposition, which does not
contain the Maxwell tensor $F^{mn}$. Note, that this term with
$\omega_{25}$ in front corresponds to the right-hand side of
Eq.~(\ref{g107}) with the multiplier $\alpha$.

In order to classify {\it nonminimal} terms in the theory of
spin-axion coupling we can use the following procedure. First,
we replace the Maxwell tensor $F^{ik}$ in (\ref{BMT2}), (\ref{d2}), and (\ref{d3}) with the tensor of nonminimal polarization-magnetization
$M^{ik}= {\cal R}^{ikmn}F_{mn}$, where the so-called nonminimal
susceptibility tensor ${\cal R}^{ikmn}$ is introduced according
to the rule
\begin{eqnarray}
&&
\hspace{-.05\textwidth}
{\cal R}^{ikmn} \equiv \frac{1}{2}q_1R\,(g^{im}g^{kn} -
g^{in}g^{km}) + q_3 R^{ikmn}+
\nonumber\\&& \hspace{-.04\textwidth}
+\frac{1}{2}q_2(R^{im}g^{kn} - R^{in}g^{km} + R^{kn}g^{im}
-R^{km}g^{in})\,. \label{sus}
\end{eqnarray}
Here $R^{mn}$ is the Ricci tensor, $R$ is the Ricci scalar, $q_1$,
$q_2$, $q_3$ are the nonminimal coupling constants (see, e.g.,
\cite{NM1,NM2,NM3,NM4} for details). Similarly, we replace the dual
Maxwell tensor $F^{*ik}$ with the tensor ${\cal M}^{ik}$:
\begin{equation}
{\cal M}^{ik} = \left[\alpha_1 \ ^{*}{\cal R}^{ikmn} + \alpha_2
{\cal R}^{*ikmn}\right]F_{mn} \,, \label{d4}
\end{equation}
where $^{*}{\cal R}^{ikmn}$ and ${\cal R}^{*ikmn}$ are the left-dual and
right-dual tensors of nonminimal susceptibility, respectively. The corresponding
coupling parameters $\omega_{ab}$ should be replaced with
$\tilde{\omega}_{ab}$. Concerning the last term in
(\ref{d3}), which does not contain the Maxwell tensor,
our strategy is to add the following nonminimal analogs:
\begin{eqnarray}
\omega^{ik}_{({\rm NM})} &=&
\nabla_l \phi
\left\{\tilde{\omega}_{25} \left[\tilde{\alpha}_1 \ ^{*}{\cal
R}^{iklj} {+} \tilde{\alpha}_2 {\cal R}^{*iklj}\right] p_j \,+
\right.
\nonumber \\ &&
\left.
+ \tilde{\omega}_{26} p^l \epsilon^{ik}_{\ \ mn} {\cal
R}^{mq}p^n p_q  \right\} . \label{d5}
\end{eqnarray}
Here ${\cal R}^{mq}={\cal R}^{mnqs}g_{ns}$. As a result of the
described procedure, we deal with a large number of
phenomenologically introduced coupling parameters. As a first step,
below we restrict ourselves by the minimal model, i.e., by the
case when $\tilde{\omega}_{ab}=0$; the nonminimal model contains a
lot of specific details and will be discussed in a special paper.

\subsubsection{Dynamic equations}\label{sec - dynamic eqs}

Following  the representation of the total tensor
$\omega^{ik}=\omega^{ik}_{(0)}+\omega^{ik}_{(1)}+\omega^{ik}_{(2)}$,
described above, we obtain the minimal dynamic equation in the
form
\begin{equation}
\frac{D p^{i}}{D\tau} = \frac{e}{m}\left\{F^{ik}p_k + F^{*ik}p_k \left[\lambda \phi
 +\mu \left(p^l \nabla_l \phi \right) \right]
\right\} . \label{211}
\end{equation}
The first term in the right-hand side of this equation is the
usual Lorentz force. The second term can be interpreted as an
axionic analog of Lorentz force, in which the Maxwell tensor
$F^{ik}$ is replaced by its dual tensor $F^{*ik}$, and the
electric charge $e$ is replaced by an effective pseudo-charge $e
\phi$. If $\lambda = 0$ identically, the model is characterized by
vanishing pseudo-charge. The coupling constant $\mu$ introduces a
completely new term, which contains  the gradient four-vector of
an effective pseudo-charge $e \phi$. We faced with similar
situation in the $SU(2)$ and $SU(3)$-symmetric gauge models, in
which the isospin and color charges, respectively, were considered
as functions, and their derivatives also entered the master equations \cite{Wong,GW1}.

The evolutionary equation for the spin with the redefinitions,
mentioned above, takes the form
\begin{eqnarray}
\frac{D S^{i}}{D\tau} &=& \frac{e}{2m} \left[g F^{ik} S_k +
\frac{(g - 2)}{m^2}  p^i F_{kl} S^k p^l \right]+
\nonumber\\ && \hspace{-.07\textwidth}
+\frac{e \lambda}{2m} \phi \left[g_{A} F^{*ik} S_k  +
\frac{(g_{A} - 2)}{m^2} p^i F^*_{kl} S^k p^l \right] +
\nonumber\\ && \hspace{-.07\textwidth}
+ \frac{e \mu}{2m} (p^l \nabla_l \phi) \left[g_{G} F^{*ik} S_k  + \frac{(g_{G} -
2)}{m^2} p^i F^*_{kl} S^k p^l \right]+
\nonumber\\ && \hspace{-.07\textwidth}
+ \frac{e\mu}{2m}(p^l \nabla_l \phi)\omega_{23} \epsilon^{ikmn}S_k p_n p^j F_{mj} +
\nonumber\\ && \hspace{-.07\textwidth}
+ \frac{\omega_{24}}{m} \nabla^l \phi \epsilon^{ikmn} S_k p_n
F_{ml}+ \frac{\omega_{25}}{m} \nabla_l \phi \epsilon^{ikln} S_k
p_n  \,. \label{51}
\end{eqnarray}
The first term in the right-hand side of this equation is the
usual term attributed to the Bargmann-Michel-Telegdi model with
anomalous magnetic moment; other terms can be indicated as its
axionic analogs. Only one term with $\omega_{25}$ in front does
not contain the Maxwell tensor thus describing the direct
spin-axion coupling.

\subsubsection{Our further strategy}\label{sec - further strategy}

We have established the phenomenological model, in which three channels of
spin-axion coupling can be distinguished. The first channel is
direct: it works even if there are no electromagnetic fields
($F_{mn}{=}0$), and the tidal  (nonminimal) interactions are absent
(${\cal R}^{ikmn}{=}0$). The second channel is indirect, the
corresponding spin-axion coupling is mediated by the Maxwell field
($F_{mn} \neq 0$, ${\cal R}^{ikmn}{=}0$). The third channel is also
indirect, and it can be opened when the model is nonminimal
(${\cal R}^{ikmn}\neq0$).
Below we put $F_{mn}{=}0$ in (\ref{51}) and consider effects of a direct spin-axion interaction only.
Next papers will be devoted to a systematic study of effects mediated
by electromagnetic fields of various structures. In the future
we also intend to add nonminimal couplings to the direct and indirect
models of spin-axion coupling.

\section{Applications of the model with direct spin-axion coupling}\label{sec - applications}

We focus now on the direct spin-axion  interactions, i.e., we
assume that $F_{mn}=0$ and ${\cal R}^{ikmn}=0$, and the gradient
four-vector $\nabla_i \phi$ is non-vanishing due to the coupling
of axions to the gravity field. In this case Eqs.~(\ref{211}) and
(\ref{51}) obtain the simple form
\begin{equation}
\frac{D p^{i}}{D\tau} = 0 \,, \qquad \frac{D S^{i}}{D\tau} =
\frac{\omega_{25}}{m} \nabla_l \phi \epsilon^{ikln} S_k  p_n  \,.\label{dsac21151}
\end{equation}
Below we study three examples. The first one relates to the
cosmological context, and the gradient four-vector $\nabla_i \phi$
is time-like (i.e., $\nabla_i \phi \nabla^i \phi>0$). The second
example relates to the case with the space-like  gradient four-vector
(i.e., $\nabla_i \phi \nabla^i \phi<0$), which can be realized,
e.g., in a spherically symmetric static space-time. The third
example corresponds to the case $\nabla_i \phi \nabla^i \phi=0$,
which can be realized in space-times with plane-wave symmetry
(gravitational waves). In all three cases we consider the spin
particle as a test one. It moves in a given space-time through
the pseudoscalar field,  which obeys the equation
\begin{equation}
\nabla^m \nabla_m \phi + {\cal V}^{\prime}(\phi^2)\phi = 0\,.
\label{1A2}
\end{equation}
Here ${\cal V}(\phi^2)$ is the potential of the pseudoscalar (axion) field.
This equation is derived from the axionic part of the total
Lagrangian
\begin{equation}
S_{({\rm axion})} = \int d^4 x \sqrt{{-}g} \ \frac{\Psi^2_0}{2} \left[{\cal V}(\phi^2) - \nabla_k \phi \nabla^k \phi \right] ,
\label{actmin}
\end{equation}
where the constant $\Psi_0$ is reciprocal to the coupling constant of the axion-photon interaction $\rho_{A \gamma \gamma}$,
i.e., $\frac{1}{\Psi_0}=\rho_{A \gamma \gamma}$ (see, e.g., \cite{NM4}). Since
the direct spin-axion effect appears if and only if the gradient
four-vector $\nabla_k \phi$ is non-vanishing, i.e., $\nabla_k \phi
\neq 0$, we assume that just the gravitational field produces the
inhomogeneity or non-stationarity of the axionic field $\phi$.

\subsection{Spin coupling to relic dark matter axions}\label{sec - relic DM}

Dark matter hypothetically contains relic axions born in the
Early Universe, and in the cosmological context the pseudoscalar
(axion) field $\phi$ can be considered   as a function of
cosmological time only, $\phi(t)$. The time variable $t$
corresponds to the following choice of the background space-time
metric:
\begin{equation}
ds^2 = dt^2 - a^2(t)\left[{dx^1}^2 + {dx^2}^2 + {dx^3}^2 \right] .
\label{1A1}
\end{equation}
In this model the gradient four-vector $\nabla_i
\phi$ is of the form $\nabla_i \phi = U_i \dot{\phi}$, where $U_i
\equiv \delta^0_i$ is the global velocity four-vector. In the context of
cosmological application we consider the
axion field potential ${\cal V}(\phi^2)$ to be of the form
${\cal V}(\phi^2)= m^2_{({\rm a})}\phi^2$, where $m_{({\rm a})}$ is the axion mass.
Then the equation of the axion field evolution (\ref{1A2}) is
\begin{equation}
\ddot{\phi} + 3\frac{\dot{a}}{a} \dot{\phi} + m^2_{({\rm a})} \phi
= 0 \,, \label{1A3}
\end{equation}
where the dot denotes a derivative with respect to time.

We
consider the space-time background to be fixed by the
corresponding gravity field equations,  and the scale factor
$a(t)$ is a known function of time \cite{S}. For instance, in the de
Sitter-type regime of cosmological expansion the scale factor is
of the form $a(t)=a(t_0)\exp{[H_0 (t{-}t_0)]}$ with the constant
Hubble function $H(t) \equiv \frac{\dot{a}}{a}=H_0$. When
$m_{({\rm a})}>\frac32 H_0$ the solution to Eq.~(\ref{1A3}) is
\begin{eqnarray}
\hspace{-.03\textwidth}
\phi(t) &=&
e^{-\frac32 H_0 (t{-}t_0)} \left\{\phi(t_0)
\cos{\Omega_{({\rm a})}}(t{-}t_0)+ \right.
\nonumber\\&& \hspace{-.05\textwidth}
\left.
+ \frac{1}{\ \ \Omega_{({\rm a})}} \left[\dot{\phi}(t_0)+
\frac32 H_0 \phi(t_0)\right]\sin{\Omega_{({\rm a})}(t{-}t_0)}
\right\} , \label{1A4}
\end{eqnarray}
where the effective axionic frequency is introduced as
\begin{equation}
\Omega_{({\rm a})} \equiv \sqrt{m^2_{({\rm a})}-\frac94 H^2_0}\,.
\label{1A5}
\end{equation}

When the electromagnetic field is absent, the equations of particle dynamics (\ref{dsac21151}) for the
metric (\ref{1A1}) are reduced to
\begin{equation}
\frac{d p_j}{d\tau} =  \frac{1}{2m} \delta_j^0 p^k p^l \dot{g}_{kl}    \,.
\label{1A6}
\end{equation}
The solution is known to be the following:
\begin{eqnarray}
&\displaystyle
p_1(t)=p_1(t_0)\,, \quad  p_2(t)=p_2(t_0) \,, \quad
p_3(t)=p_3(t_0) \,,
&\nonumber\\ &\displaystyle
p_0(t) = \sqrt{m^2 + \frac{q^2}{a^2(t)}}\,, \label{1A7}
&
\end{eqnarray}
where $q^2 \equiv p^2_1(t_0){+}p^2_2(t_0){+}p^2_3(t_0)$ is the
constant quantity. Since the space-time is spatially  isotropic,
we assume that the particle had only one non-vanishing component
at $t=t_0$, say, $p_3(t_0)\neq 0$, and consider (for the
illustration) the following initial data:
\begin{equation}
p_1(t_0) = 0 \,, \quad  p_2(t_0) =0 \,, \quad S^3(t_0) = 0 \,.
\label{1A700}
\end{equation}
Clearly, for this case at an arbitrary time moment $p_1(t)=p_2(t)=0$
and $p_3(t)= p_3(t_0)$, so that the cosmological time $t$ and the
proper time  $\tau$ along the particle world-line, are linked
by the relationship
\begin{equation}
\tau =\int_{t_0}^t \frac{m a(t) \ dt}{\sqrt{m^2 a^2(t) +  q^2}}
\,. \label{2A700}
\end{equation}
The equation of spin dynamics (\ref{dsac21151}) is now of the form
\begin{equation}
\frac{D S^{i}}{D\tau} =  \frac{\omega_{25}}{m} \epsilon^{ik03} S_k
p_3 \dot{\phi} \,.\label{1A8}
\end{equation}
Its right-hand side does not equal to zero only for two values of
indices: $i{=}1$ and $i{=}2$. This means that using the property
$S^k p_k {=} S^0p_0{+}S^3p_3{=}0$ we can write the equations for
the components $S^{3}$ and $S^{0}$ as follows:
\begin{equation}
\frac{d S^{3}}{d t} + H_0 S^3 \left(1+\frac{p^2_3}{a^2 p_0^2}
\right)= 0 \,, \quad S^{0} = - \frac{S^{3} p_3}{p_0}\,.
\label{31A8}
\end{equation}
For the initial value $S^3(t_0)=0$ these equations have the only
trivial solution $S^3(t)=0$, $S^0(t)=0$. In other words the
transverse spin components only are influenced by the axion
environment and the longitudinal component is not touched.

From the normalization condition $S^k S_k=-{\cal S}^2$ we
obtain $a^{-2}(t)(S^2_1+S^2_2)={\cal S}^2$, which is the hint to
introduce two convenient variables
\begin{equation}
S_{+}(t)= \frac{S_1}{a(t)} \,, \quad S_{-}(t)= \frac{S_2}{a(t)}
\,, \label{1A9}
\end{equation}
so that
\begin{equation}
S^2_{+} + S^2_{-}={\cal S}^2 =\mbox{const} \,.
\label{1A9norm}
\end{equation}
In these terms the equations for the transverse components take
the form
\begin{equation}
\dot{S}_{+} = - \Omega(t) S_{-} \,, \quad \dot{S}_{-} = \Omega(t)
S_{+} \,, \label{1A10}
\end{equation}
where
\begin{equation}
\Omega(t)=  \frac{\omega_{25} p_3(t_0) \  \dot{\phi}(t)}{a(t)p^0(t)} \,.
\label{1A11}
\end{equation}
The solutions to (\ref{1A10}) are of harmonic type
\begin{equation}
S_{+} = {\cal S} \cos{\Psi(t)} \,,
\quad  S_{-} = {\cal S} \sin{\Psi(t)} \,,
\label{1A12}
\end{equation}
where the phase of rotation is presented as a formal integral
\begin{equation}
\Psi(t)= \int_{t_0}^t \Omega(t)dt + \Psi(t_0)  \,.
\label{1A129}
\end{equation}
For the illustration of the obtained exact solution we assume, first, that
the particle is ultrarelativistic ($q^2>>m^2 a^2(t)$) and $p_3(t_0)$ is positive;
second, we choose the time moment $t_0$ so that $\dot{\phi}(t_0)=0$;
third, we put  for simplicity $\Psi(t_0){=}\omega_{25}\phi(t_0)$.
Then, keeping in mind that nowadays $m_{({\rm a})}\gg H_0$, we find the phase $\Psi(t)$ in the explicit form (see (\ref{1A4}))
\begin{equation}
\Psi(t)=  \sigma \cos{m_{({\rm a})}(t-t_0)}\,, \quad \sigma \equiv \omega_{25} \phi(t_0)\,.
\label{1A13}
\end{equation}
The solutions for the spin components can be now presented as follows
\begin{eqnarray}
&&\displaystyle\hspace{-.07\textwidth}
S_{+} = {\cal S} \cos\{ \sigma \cos{[m_{({\rm a})}(t-t_0)}] \} =
\nonumber\\ &&\displaystyle\hspace{-.05\textwidth}
=J_0(\sigma) {+} 2 \sum_{n{=}1}^{\infty} ({-}1)^n J_{2n}(\sigma) \cos{[2n m_{({\rm a})}(t{-}t_0)]} \,,
\label{Bessel1}
\\ &&\displaystyle\hspace{-.07\textwidth}
S_{-} = {\cal S} \sin\{\sigma \cos{m_{({\rm a})}(t-t_0)}]\} =
\nonumber\\ &&\displaystyle\hspace{-.05\textwidth}
= 2 \sum_{n{=}0}^{\infty} ({-}1)^n J_{2n{+}1}(\sigma) \cos{[(2n{+}1) m_{({\rm a})}(t{-}t_0)]} \,,
\label{Bessel2}
\end{eqnarray}
where $J_{m}(\sigma)$ are the Bessel functions of the first kind.
We deal with sophisticated spin precession, for which the phase of precession $\Psi(t)$
is the harmonic function oscillating with the axionic frequency $\Omega_{({\rm a})}{=}m_{({\rm a})}$ and the amplitude $\sigma = \omega_{25}\phi(t_0)$.


\subsection{Spin precession of relativistic charged particle in storage rings}
\label{sec - StorageRing}

When a relativistic charged spin particle moves in the constant magnetic field $B$, the motion is known to be circular, and the quantity reciprocal to the Larmor frequency, $\omega^{-1}_{{\rm L}}=\frac{m}{eB}$, predetermines the time scale of dynamic processes. In fact, for such motion the cosmological phenomena can be considered as extremely slow, and one can put $H(t)\to 0$. The scale factor $a(t)$ can be replaced by constant $a(t_0)$ and absorbed into the redefined coordinates (in fact, one can put $a(t) \to 1$). Since the size of the storage ring is much smaller than the typical size of the dark matter inhomogeneity, we can neglect a spatial dependence and consider the axion field as a function of time only.  In our case the quantity $\dot{\phi}$ can be expressed in terms of the energy density scalar 
$W_{({\rm a})}$ and pressure $P_{({\rm a})}$ attributed to the axionic dark
matter as follows (see, e.g., \cite{P1}):
\begin{equation}
\dot{\phi} = \pm \frac{1}{\Psi_0} \sqrt{W_{({\rm a})}(t)+P_{({\rm
a})}(t)} \,. \label{cl18}
\end{equation}
When the axionic dark matter is cold,
i.e., $P_{({\rm a})}{=}0$, and $W_{({\rm a})}{=}\rho_{({\rm a})}$,
where $\rho_{({\rm a})}$ is the mass density of the dark matter,
this formula is simplified, respectively, as
\begin{equation}
\dot{\phi} = \pm \frac{1}{\Psi_0} \sqrt{\rho_{({\rm a})}(t)} \,. \label{cl18m}
\end{equation}
Also, for the sake of simplicity, we can neglect the deformation of the initial magnetic field by the axionic field, and put equal to zero all the new coupling constants except $\omega_{25}$.

Let the magnetic field be directed along the $x^3\equiv z$ axis, i.e., only one component of the Maxwell tensor, $F_{12}={\rm const}$, is non-vanishing.
In the cylindrical coordinates $\{\rho, \varphi, z\}$ with the metric
\begin{equation}
ds^2 = dt^2 - (dz^2 + {d\rho}^2 + \rho^2 {d\varphi}^2 )
\label{h77}
\end{equation}
the equation of particle motion
\begin{equation}
\frac{Dp_j}{D\tau} = \frac{e}{m} F_{jk} p^k   \,,
\label{h1}
\end{equation}
can be transformed as
\begin{equation}
\frac{dp_j}{dt} = - \frac{\rho}{p_0} \delta_j^{\rho} {p^{\varphi}}^2 + \frac{e}{p_0} F_{jk} p^k  \,,
\label{h81}
\end{equation}
and gives the evident solutions
\begin{eqnarray}
&&\!\!\!\!\!\!\!\!\!\!\!\!\!\!\!
p_z(t)=p_z(0)=0 \,, \quad  p_{\rho}(t)=0 \,,
\label{h199}
\\&&\!\!\!\!\!\!\!\!\!\!\!\!\!\!\!
p_{\varphi} = -e \rho F_{\rho \varphi} = - e \rho^2 F_{12} \,,
\label{h197}
\\&&\!\!\!\!\!\!\!\!\!\!\!\!\!\!\!
\rho(t) = R = {\rm const} \,, \quad \varphi(t) = \Omega_{({\rm B})} t \,,
\label{h196}
\\&&\!\!\!\!\!\!\!\!\!\!\!\!\!\!\!
p_0 = \sqrt{m^2{+} e^2 R^2 F^2_{12}}\,, \quad t= \tau \sqrt{1{+} \frac{e^2 R^2 F^2_{12}}{m^2}} \,.
\label{h195}
\end{eqnarray}
Here $R$ is the radius of the circular orbit, and the quantity $\Omega_{({\rm B})}$ given by
\begin{equation}
\Omega_{({\rm B})} = \frac{eF_{12}}{p_0} = {\rm const} \,,
\label{h4}
\end{equation}
is the relativistic angular frequency of rotation (the relativistic Larmor frequency).

The equations for the spin evolution
\begin{equation}
\frac{DS^i}{D\tau} = \frac{\omega_{25} \dot{\phi}}{m} \epsilon^{ik0n} S_k p_n + \frac{e}{m} F^{i}_{\ k} S^k   \label{h5}
\end{equation}
rewritten as
\begin{equation}
p_0\frac{dS^i}{dt} {+} \frac{p_{\varphi}}{\rho}\left(g^{i\varphi} S^{\rho}{-}g^{i\rho} S^{\varphi} \right)=  \frac{\omega_{25} \dot{\phi}}{\rho}E^{ik0\varphi} S_k p_{\varphi} {+} e F^{i}_{\ k} S^k   \label{h58}
\end{equation}
give
\begin{eqnarray}
&&
\dot{S}^{0} = 0 \,, \quad \dot{S}^{\varphi} = 0\,,  \label{h7a}
\\&&
\dot{S}^z =  \omega_{25} \dot{\phi} \Omega_{({\rm B})} R \ S^{\rho} \,,  \label{h7b}
\\&&
\dot{S}^{\rho} = - \omega_{25} \dot{\phi} \Omega_{({\rm B})} R \ S^z  \,.  \label{h7c}
\end{eqnarray}
Physically motivated solutions to these equations are
\begin{eqnarray}
&
S^{0}(t) = 0 \,, \quad S^{\varphi}(t) = 0\,,  \label{h71}
&\\&
S^z(t) = - {\cal S} \cos{\Psi_{({\rm H})}(t)} \,,
&\\&
S^{\rho}(t) = {\cal S} \sin{\Psi_{({\rm H})}(t)} \,,
\label{hy7}
&
\end{eqnarray}
where the hybrid precession phase $\Psi_{({\rm H})}$ is given by
\begin{equation}
\Psi_{({\rm H})}(t) {=} \Psi(t) \  \Omega_{({\rm B})} R = \Psi(t) \frac{eF_{12} R}{\sqrt{m^2{+} e^2 R^2 F^2_{12}}}  \,,  \label{hp7}
\end{equation}
and $\Psi(t)$ is the axionic phase (we put here $t_0=0$)
\begin{eqnarray}
&\displaystyle
\Psi(t) = \omega_{25}[\phi(t)-\phi(0)] =
&\nonumber\\ &\displaystyle
= \omega_{25} \left[\phi(0) (\cos{\Omega_{({\rm a})}t}{-}1){+} \frac{\dot{\phi}(0)}{\ \ \Omega_{({\rm a})}} \sin{\Omega_{({\rm a})}t} \right] \simeq
&\nonumber\\ &\displaystyle
\simeq \omega_{25} \  \dot{\phi}(0) \ t \,.  
\label{hp77}
&
\end{eqnarray}
Clearly, the spin four-vector is orthogonal to the particle momentum four-vector, i.e., $p_kS^k=0$.
This solution can be illustrated as follows. If $\omega_{25}\dot{\phi}=0$, the particle has the spin three-vector directed along the magnetic field, and this direction is conserved during the particle circular motion. If $\omega_{25}\dot{\phi} \neq 0$ the spins start to precess in the plane $\rho O z$ according the law described by formulas (\ref{h71})-(\ref{hy7}); the frequency of the precession depends on the particle energy, or, equivalently, on the orbit radius $R$.
From the experimental point of view, if the polarized beam of electrons is formed in a storage ring, and all the spins are initially directed perpendicularly to the ring plane, one can expect, that the axionically induced spin rotation will start, and the distribution of the angles between the spins and the ring plane will be a predicted function of time and the particle energy.  
In the ultrarelativistic regime, when $m\to 0$ effectively, one obtains from (\ref{hp7}) that
$\Psi_{({\rm H})}(t) \to \Psi(t)$ and the dependence on the parameter $R$ disappears (some specific details of ultrarelativistic spin particle motion can be found also in \cite{ultra}).


\subsection{Spin dynamics in the field of an axion star}\label{sec - axion star}

In this application we consider a spherically symmetric static
axionically active object, which is characterized by the metric
\begin{equation}
ds^2 = B(r) dt^2 -A(r) dr^2 - r^2 \left(d \theta^2 +
\sin^2{\theta} d \varphi^2 \right) \,. \label{as1}
\end{equation}
The axion field is assumed to depend on the radial coordinate $r$
only. The gradient four-vector $\nabla_i \phi = \delta_i^r
\phi^{\prime}(r)$ (the prime denotes the derivative with respect
to $r$) is now space-like, i.e., $\nabla_i \phi \nabla^i \phi = -
\frac{1}{A} {\phi^{\prime}}^2<0$.
When we consider the distribution of the pseudoscalar (axion) field in this static model, and the gravity field is assumed to be strong,
it seems to be reasonable to use the extended potential
\begin{equation}
{\cal V}(\phi^2) = m^2_{({\rm a})} \phi^2 + \frac12 \nu_{({\rm a})} (\phi^2-\phi^2_{*})^2  \,. \label{phi4}
\end{equation}
The function $\phi(r)$ satisfies now the equation
\begin{eqnarray}
&& \hspace{-.07\textwidth}
 \phi^{\prime\prime} +
 \phi^{\prime}\left[\frac12 \left(\frac{B^{\prime}}{B}-\frac{A^{\prime}}{A} \right)
 + \frac{2}{r}\right] =
\nonumber\\ && \hspace{-.05\textwidth}
= A(r)\phi \left[m^2_{({\rm a})} + \nu_{({\rm a})} (\phi^2-\phi^2_{*}) \right], \label{as2}
\end{eqnarray}
which is (due to the absence of electromagnetic field) formally the same as for the gravitating
scalar field $\Phi(r)$ (see, e.g., \cite{Scal1,Scal2,Scal3}).
There is no need to present explicit solutions of (\ref{as2}) for our purposes. Examples of solutions with
asymptotically flat space-time and scalar fields vanishing at $r\to \infty$ can be found in \cite{Scal1,Scal2,Scal3}.

In the static spherically symmetric case the plane of the particle
motion can be chosen as the equatorial one $\theta=\frac{\pi}{2}$,
and the first equation in (\ref{dsac21151}) is known to give four integrals of
motion \cite{GraCos}:
\begin{eqnarray}
&\displaystyle
 p_0 {=} K {=} \mbox{const} \,, \quad p_{\theta}=0 \,, \quad  p_{\varphi} = {-} J =\mbox{const} \,,   \label{as3}
 &\displaystyle\\
&\displaystyle
p^{r} = \frac{1}{\sqrt{A}}\sqrt{\frac{K^2}{B} - \frac{J^2}{r^2} - E^2}
\,, \quad E = \mbox{const}\,. \label{as4}
&\displaystyle
\end{eqnarray}
For the asymptotically flat space-time with
$A(\infty){=}1$ and $B(\infty){=}1$ the normalization condition $p_ip^i {=}m^2$
yields
\begin{equation}
K = p_0 (\infty) = \sqrt{m^2{+}{p_{r}}^2(\infty)} \,, \label{as5}
\end{equation}
i.e., the constant $K$ usually describes the particle energy at
infinity. The constant $J$ relates to the conserved particle
angular momentum. The constant $E$ regulates the parameter along
the particle world-line, e.g., when $d\tau = ds$, we see that
$E=m$.

In order to analyze properly the equations of spin evolution
(\ref{dsac21151}) we distinguish two specific types of particle motion:
first, the radial motion; second, the motion along a circular orbit.

\subsubsection{Radial particle motion}\label{sec - radial}

For this type of motion $p_{\theta}{=}0$, $p_{\varphi}{=}0$ and the equations of spin evolution
\begin{equation}
\frac{D S^{i}}{D\tau} =
\frac{\omega_{25}}{m} \phi^{\prime}(r) \  \epsilon^{ikr0} S_k  p_0
\label{nn}
\end{equation}
can be split into two independent subsets. The
first subset contains the components $S^0$ and $S^r$ only, and
does not include the coupling term proportional to $\omega_{25}$.
Keeping in mind that the condition $S^k p_k=0$ leads to
$S^0p_0+S^r p_r=0$, we can write this subset of equations as
follows:
\begin{eqnarray}
&\displaystyle
\frac{d S^0}{d \tau} + S^0 \ \left\{\frac{B^{\prime}(r)}{2mAB p^r}\left[2\frac{K^2}{B(r)}-m^2 \right]\right\}=0 \,,
\label{as7}
&\displaystyle\\
&\displaystyle
\frac{d S^r}{d \tau} + S^r \ \left\{\frac{ p^r}{2m}\left(\frac{A^{\prime}}{A}+\frac{B^{\prime}}{B}\right)\right\}=0 \,.
\label{as71}
&
\end{eqnarray}
As in the first application we assume that $S^r(0)=0$, providing
that $S^0(0)=0$ from the orthogonality condition. Then the
solutions to the equations (\ref{as7}) and (\ref{as71}) are
trivial $S^0(\tau)=0$ and $S^r(\tau)=0$.

The second subset of (\ref{nn})
\begin{eqnarray}
&\displaystyle
\frac{d S^{\theta}}{d\tau}+ S^{\theta} \ \left(\frac{p^r}{mr}\right)=
-\frac{\omega_{25}\phi^{\prime}}{m \sqrt{AB}r^2} p_0 S_{\varphi}
\,,
\label{as8}
&\displaystyle\\
&\displaystyle
\frac{d S^{\varphi}}{d\tau}+ S^{\varphi} \ \left(\frac{p^r}{mr}\right)=\frac{\omega_{25}\phi^{\prime}}{m \sqrt{AB}r^2}
p_0 S_\theta \,,\label{as81}
&
\end{eqnarray}
can be transformed into
\begin{equation}
\frac{d S_{+}}{dr} = - \Omega(r) S_{-} \,, \quad \frac{d S_{-}}{dr} = \Omega(r) S_{+} \,,
\label{as9}
\end{equation}
using the relation between $\tau$ and $r$ (see (\ref{as4}))
\begin{equation}
\tau = \int_{\infty}^r dr \ \frac{m \sqrt{AB(r)}}{\sqrt{K^2-E^2 B(r)}} \,, \label{as10}
\end{equation}
and the following definitions:
\begin{eqnarray}
&\displaystyle
S_{+} \equiv r S^{\theta} \,, \quad  S_{-} \equiv r S^{\varphi} \,,
\label{as11}
&\displaystyle\\
&\displaystyle
\Omega(r) \equiv \frac{K \omega_{25}\phi^{\prime}(r)}{\sqrt{K^2- E^2 B(r)}}
\,. \label{as111}
&
\end{eqnarray}
Clearly, the solutions to (\ref{as9}) are
\begin{eqnarray}
&\displaystyle
S_{+} (r) =  {\cal S} \cos{\Psi(r)} \,, \quad  S_{-} = {\cal S} \sin{\Psi(r)} \,,
\label{as12}
&\displaystyle\\
&\displaystyle
\Psi(r) = \Psi(\infty) + \int_{\infty}^r dr  \Omega(r) \,. \label{as1212}
&
\end{eqnarray}
Thus, when the particle moves in the radial direction, one
deals with a spin turn in the plane $(\theta,\phi)$; the quantity
$\Omega(r)$ plays the role of the rate of turn with respect to
radial variable $r$, and $\Psi(r)$ describes the cumulative angle
of turn.

\subsubsection{Circular particle motion}\label{sec - circular}

The circular motion is characterized by $r=R=$const, so that
$p^r=0$ and thus
\begin{equation}
\frac{K^2}{B(R)} - \frac{J^2}{R^2} = E^2
\,. \label{as47}
\end{equation}
We consider a stable orbit and thus the radial component of the
gravitational force should vanish on the orbit, providing
$\frac{Dp^r}{D\tau} = 0$, or
\begin{equation}
\frac{B^{\prime}(R)}{B^2(R)} = \frac{2J^2}{K^2 R^3} \,.
\label{as48}
\end{equation}
Also, as previously, we have that $p_0 = K$, $p_{\theta}=0$,
$p_{\varphi} = - J$, but now the particle can not reach infinity,
and we have to re-define the constant $K$. For instance, we obtain
from (\ref{as47}) and (\ref{as48}) that
\begin{equation}
J^2 = \frac{E^2 R^3 B^{\prime}(R)}{2B- RB^{\prime}(R)} \,, \quad K
= E \sqrt{\frac{2 B^2(R)}{2B- RB^{\prime}(R)}} \,. \label{7as47}
\end{equation}
The first equation of (\ref{7as47}) gives implicitly the radius of
the circular orbit as a function of orbital moment, $R(J)$. The
second equation defines the energy of particle on the given orbit
as a function of obtained radius, $K(R)$. We assume that the
inequality $2B(R)>RB^{\prime}(R)$ is satisfied (for instance, for
the Schwarzschild metric $B(r)=1-\frac{2GM}{r}$ this inequality
means that $R>\frac32 r_g=3GM$).

The orthogonality condition $S^k p_k=0$ and the integrals of motion
(\ref{as3}) provide that the components
$S^0$ and $S^{\varphi}$ are proportional one to another, $S^0 =
S^{\varphi} \left( \frac{J}{K}\right)$. For the circular orbit one
can replace the differentiation with respect to proper time $\tau$
with the azimuthal angle $\varphi$ due to the relationship
$\frac{d\varphi}{ds}{=}\frac{p^{\varphi}}{m}{=} \frac{J}{m R^2}$
(recall, that in the case $s=\tau$ we obtain $E=m$).
With this replacement three independent equations for spin
evolution take the following form:
\begin{eqnarray}
&\displaystyle
\frac{d S^r}{d\varphi} = S^{\varphi} \cdot H^{r}(R) \,,
\quad \frac{d S^{\theta}}{d\varphi} = S^{\varphi} \cdot H^{\theta}(R) \,,
&\nonumber\\
&\displaystyle
\frac{d S^{\varphi}}{d\varphi} +  \frac{S^{r}}{R} = - S^{\theta} \cdot H^{\varphi}(R) \,.
\label{ss1}
&
\end{eqnarray}
Here, for short, we introduced three auxiliary functions of the
radius $R$
\begin{eqnarray}
&\displaystyle
H^{r}(R) = \frac{RE^2 B(R)}{K^2 A(R)} \,, \quad H^{\varphi}(R) = H^{\theta}(R) \frac{K^2}{E^2 B(R)} \,,
&\nonumber \\
&\displaystyle
H^{\theta}(R) = \frac{\omega_{25} R^2 E^2 \sqrt{B(R)} \phi^{\prime}(R)}{K J\sqrt{A(R)}}
\,.
\label{ss2}
&
\end{eqnarray}
The evident differential consequence of (\ref{ss1})-(\ref{ss2}) is the following equation of the second order for $S^{\varphi}$:
\begin{equation}
\frac{d^2 S^{\varphi}}{d\varphi^2} + I^2(R) \ S^{\varphi} = 0 \,,
\label{ss4}
\end{equation}
where
\begin{equation}
I^2(R) = \frac{1}{R} H^{r} + H^{\theta} H^{\phi} =
\frac{E^2 B}{K^2 A} {+} \frac{\omega^2_{25} R^4 E^2 {\phi^{\prime}}^2}{J^2 A} \,.
\label{ss9}
\end{equation}
For the positive metric functions $B(R)>0$ and $A(R)>0$
the solution to this equation is harmonic function of the
azimuthal angle
\begin{equation}
S^{\varphi}(\varphi) = C_1 \cos{I\varphi} + C_2 \sin{I \varphi} \,,
\label{ss5}
\end{equation}
providing the following solutions in terms of $\tau$:
\begin{eqnarray}
&\displaystyle
S^{\varphi}(\tau) = C_1 \cos{\Omega \tau} +  C_2 \sin{\Omega \tau} \,,
&\displaystyle\nonumber\\
&\displaystyle
S^{0}(\tau) = \frac{J}{K} \left[C_1\cos{\Omega \tau} +  C_2 \sin{\Omega \tau}\right] \,.
\label{ss51}
&
\end{eqnarray}
The quantity $\Omega {=} \frac{I \cdot J}{mR^2}$
plays the role of frequency of the spin turn.
Other components of the spin four-vector are, respectively:
\begin{eqnarray}
&\displaystyle
S^{r}(\tau) = \frac{H^r(R)}{I(R)}\left[C_1 \sin{\Omega \tau} -  C_2 \cos{\Omega \tau} \right] + C_3\,,
&\displaystyle\nonumber\\
&\displaystyle
S^{\theta}(\tau) = \frac{H^{\theta}(R)}{I(R)}\left[C_1 \sin{\Omega \tau} -  C_2 \cos{\Omega \tau} \right]+ C_4\,.
\label{ss7}
&
\end{eqnarray}
The constants of integration $C_1$, $C_2$, $C_3$, $C_4$ are
connected by the normalization condition
\begin{equation}
B {S^{0}}^2 - A {S^{r}}^2 -r^2 {S^{\theta}}^2 - r^2 {S^{\varphi}}^2= - {\cal S}^2 = \mbox{const}\,,
\label{ss11}
\end{equation}
which at $r=R$ yields two relationships
\begin{eqnarray}
&\displaystyle\hspace{-.03\textwidth}
{\cal S}^2 = \frac{BE^2 R^2}{K^2} \left(C_1^2{+}C_2^2\right) {+}
R^2 C_4^2 \left[1{+} \frac{\omega^2_{25} R^4 K^2 {\phi^{\prime}}^2}{J^2 B} \right],
&\nonumber\\
&\displaystyle\hspace{-.03\textwidth}
C_3 = - C_4 \ \frac{\omega_{25} R^3 \phi^{\prime}(R) K}{J \sqrt{AB(R)}} \,.
\label{ss14}
&
\end{eqnarray}
On the other hand, three independent constants of integration $C_1$, $C_2$, $C_4$,
are connected with initial values $S^{\varphi}(0)$, $S^{\theta}(0)$, $S^{r}(0)$ as follows:
\begin{eqnarray}
&\displaystyle
C_1 = S^{\varphi}(0) \,, \quad
C_2 = - \frac{S^{r}(0)}{R I(R)} - S^{\theta}(0) \frac{K^2 H^{\theta}}{E^2I B(R)} \,,
&\nonumber\\
&\displaystyle
C_4 = - S^{r}(0) \frac{H^{\theta}}{RI^2} + S^{\theta}(0) \frac{H^{r}}{RI^2}
\,.
\label{ss16}
&
\end{eqnarray}
The problem is solved completely. The formulas
(\ref{ss51})-(\ref{ss7}) with constants given by (\ref{ss14}),
(\ref{ss16}) and auxiliary quantities (\ref{ss2}), (\ref{ss9})
describe the axionically induced turn of the spin four-vector of
the particle moving along the circular  orbit around the static
spherically symmetric gravitating object. When $\omega_{25}
\phi^{\prime}{=}0$, we see that $I {=}\frac{E}{K}\sqrt{\frac{B}{A}}$, and the corresponding frequency
$\Omega {=} \frac{IJ}{mR^2}$ takes the form $\Omega =
\Omega_{({\rm geodesic})} =\frac{J}{KR^2}\sqrt{\frac{B}{A}}$,
describing the geodesic precession. This fact allows us to
indicate the term $\Omega_{({\rm axion})} {=} \frac{\omega_{25}
{\phi^{\prime}}(R)}{ \sqrt{A(R)}}$ as the axionic frequency. With
this terminology we can say, that the total frequency $\Omega$
satisfies the equality
\begin{equation}
\Omega = \sqrt{\Omega^2_{({\rm geodesic})} + \Omega^2_{({\rm
axion})}} \,, \label{ss76}
\end{equation}
and can be called as a hybrid frequency of the geodesic-axionic
precession. Finally, it should be mentioned that one can introduce local
frequency $\omega$ instead of $\Omega$, using the equality $\omega
dt = \Omega d \tau$. Clearly, we obtain that $\omega=\Omega B(R)
\frac{m}{K}$.

\subsection{Spin precession induced by plane-symmetric axion-gravitational waves}\label{sec - GW}

The third application of the model relates to the case, when the
gradient four-vector  $\nabla_i \phi$ is the null one,
i.e., $\nabla_i \phi \nabla^i \phi=0$. It can be realized, e.g.,
in the model with plane-wave symmetry   \cite{MTW,G,NM4}. The
corresponding space-time metric is of the form:
\begin{equation}
ds^2=2dudv{-}L^2 \left[e^{2\beta}(dx^2)^2{+}e^{{-}2\beta}(dx^3)^2\right]
\,, \label{GW1}
\end{equation}
where $u=\frac{t{-}x^1}{\sqrt2}$ and $v=\frac{t{+}x^1}{\sqrt2}$ are the retarded and advanced times, respectively, and
two metric functions $L(u)$ and $\beta(u)$ depend on the
retarded time $u$ only. On the plane-wave front $u=0$ the initial
data are fixed in the form
\begin{equation}
 L(0)=1 \,, \quad L^{\prime}(0) =0 \,, \quad \beta(0)
=0  \,. \label{GW11}
\end{equation}
We assume that the background pseudoscalar (axion) field also
depends on retarded time only, $\phi=\phi(u)$, providing the
condition $\nabla_i \phi \nabla^i \phi=0$ automatically. Exact solutions of this
type (in particular, the solution linear in the retarded time) can
be found in \cite{NM4}.

The equations of particle dynamics in the metric (\ref{GW1}) are
known to yield (see, e.g., \cite{GW1})
\begin{eqnarray}
&\displaystyle
p^u = p_v=C_v \,, \quad p_2 = C_2 \,, \quad p_3 =C_3 \,,
&\nonumber\\
&\displaystyle
p^v=p_u = \frac{m^2 +
L^{-2}(e^{-2\beta}C_2^2+e^{2\beta}C_3^2)}{2C_v} \,. \label{GW2}
&
\end{eqnarray}
Here $C_v$, $C_2$, $C_3$ are constants of integration.

Our aim is to solve the equations of the spin evolution, which can
be reduced now to the following three independent equations:
\begin{eqnarray}
&&\hspace{-.07\textwidth}
\frac{d}{du}S^u =0 \qquad \Rightarrow \qquad S^u=S_v= \mbox{const}
\label{GW3}
\\ &&\hspace{-.07\textwidth}
\frac{e^{-\beta}}{L}\frac{d}{du}\left(L e^{\beta} S^2 \right) +
\frac{C_2}{2C_v} S_v \left(L^{-2}e^{-2\beta} \right)^{\prime}
=
\nonumber\\
&&
=\frac{\omega_{25}\phi^{\prime}}{L^2}\left(S_v
\frac{C_3}{C_v}-S_3\right) \,, \label{GW4}
\\
&&\hspace{-.07\textwidth}
\frac{e^{\beta}}{L}\frac{d}{du}\left(L e^{-\beta} S^3 \right) +
\frac{C_3}{2C_v} S_v \left(L^{-2}e^{2\beta} \right)^{\prime} =
\nonumber\\
&&
=
-\frac{\omega_{25}\phi^{\prime}}{L^2}\left(S_v
\frac{C_2}{C_v}-S_2\right) \,.\label{GW5}
\end{eqnarray}
Here we used the relationship $u=\tau \frac{C_v}{m}$ between the
parameter $\tau$ and the retarded time $u$, which is the
consequence of the equation $m\frac{du}{ds}=p^u=C_v$ (we
restrict ourselves by the case, when $u{=}0$ corresponds to $\tau {=}0$). The prime
denotes here the derivative with respect to retarded time. We present
only three equations from four, since the component $S^v{=}S_u$ can
be found from one of the two integrals
\begin{eqnarray}
&\displaystyle
2S_u S_v = \left(L e^{\beta}
S^2\right)^2+\left(Le^{-\beta}S^3\right)^2 - {\cal S}^2 \,, \label{GW6}
&\displaystyle\\
&\displaystyle
S_u C_v + S_v p_u + S^2 C_2 + S^3 C_3 =0 \,. \label{GW61}
&
\end{eqnarray}
Clearly, when $S_v{=}0$, one can extract $S_u$ from the second
relationship only ($C_v \neq 0$ for massive particles).

In order to solve the key equations (\ref{GW4}) and (\ref{GW5}),
it is convenient to use the following
auxiliary functions:
\begin{eqnarray}
&\displaystyle
S_{-}(u) = L e^{\beta} S^2 \,, \quad S_{+}(u) = L e^{-\beta} S^3
\,,
&\nonumber\\
&\displaystyle
\Omega(u) = \omega_{25} \phi^{\prime}(u) \,,
\label{GW88}
&\\
&\displaystyle
 f(u) = S_v \frac{C_2}{C_v}
\left(\frac{e^{-\beta}}{L}\right) \,, \quad g(u) = S_v
\frac{C_3}{C_v} \left(\frac{e^{\beta}}{L}\right) \,. \label{GW89}
&\nonumber
\end{eqnarray}
In these terms the equations (\ref{GW4}), (\ref{GW5}) take the
form
\begin{eqnarray}
&\displaystyle
 \frac{d}{du}S_{-}  = \Omega(u) S_{+} + \Omega g(u)
-f^{\prime}(u)  \,,
&\nonumber\\
&\displaystyle
\frac{d}{du}S_{+}  = -\Omega(u)S_{-} -
\Omega f(u) - g^{\prime}(u)  \,, \label{2GW5}
&
\end{eqnarray}
and their solutions happen to be very simple:
\begin{eqnarray}
&\displaystyle
S_{+} = {\cal A} \cos{\Psi(u)} - g(u) \,,
&\nonumber\\
&\displaystyle
S_{-}={\cal A} \sin{\Psi(u)} - f(u)\,,
&\label{GW9}
\\
&\displaystyle
\Psi(u)= \Psi(0)+  \omega_{25} [\phi(u)-\phi(0)] \,.
&\nonumber
\end{eqnarray}
Here ${\cal A}$ is an integration  constant. When the integral
$S_v$ is non-vanishing, the last unknown function $S_u(u)$ reads
\begin{equation}
S_{u} {=} \frac{1}{2S_v} \left[f^2{+} g^2 {-}
2 {\cal A} \left(g \cos{\Psi} {+} f \sin{\Psi} \right) {+} {\cal
A}^2{-}{\cal S}^2 \right] , \label{GW91}
\end{equation}
and the constant ${\cal A}$ can be found from the orthogonality
condition (\ref{GW61}) yielding
\begin{equation}
 {\cal A}=
\sqrt{{\cal S}^2 - \frac{m^2 S^2_v}{C^2_v}} \,. \label{4GW91}
\end{equation}
Below we illustrate the obtained exact solutions by the examples
of longitudinal and transversal particle motion with respect to the plane
front of the gravitational wave.

\subsubsection{Longitudinal motion}\label{sec - GW longitudinal}

Let the spinning particle start to move along the $x^1$ - axis, i.e.,
$p_2(0)=C_2=0$, $p_3(0)=C_3=0$. According to (\ref{GW2}) at $u>0$
the particle keeps the direction of motion, $p_2(u)=p_3(u)=0$.
From (\ref{GW3}) and (\ref{GW6}) it follows that the components
$S_u$ and $S_v$ do not feel the influence of axions, and we assume
that $S_u(0)=S_v(0)=0$, i.e., at $u=0$ there were only two
non-vanishing spin four-vector components, $S^2(0) \neq 0$ and
$S^3(0) \neq 0$. For such initial data we obtain immediately that
$f(u){=} 0$, $g(u){=}0$, ${\cal A}{=} {\cal S}$ and thus
\begin{eqnarray}
&\displaystyle
S_u(u) =0  \,, \quad S_v(u) =0 \,,
&\nonumber \\
&\displaystyle
S^2(u)= {\cal S} \left(\frac{e^{-\beta}}{L} \right) \sin{\Psi(u)} \,,
&\label{GW81}\\
&\displaystyle
S^3(u)= {\cal S}  \left(\frac{e^{\beta}}{L} \right) \cos{\Psi(u)} \,.
&\nonumber
\label{GW8}
\end{eqnarray}
Again we deal with axionically induced spin rotation with the
frequency $\Omega(u)=\omega_{25} \phi^{\prime}(u)$.

\subsubsection{Example of a transversal motion}\label{sec - GW transversal}

Let the particle start to move at $u{=}0$ in the direction $x^2$
(in the front plane of the gravitational wave), and have initially
only one non-vanishing component of the spin four-vector $S^3(0)
\neq 0$. Mathematically this is possible if
\begin{equation}
S_v=0 \,, \quad \Psi(0){=}0 \  \Rightarrow \  f(u){=}g(u){=}0 \,,
\quad {\cal A}={\cal S} \,. \label{8GW8}
\end{equation}
Then the exact solution obtained above gives:
\begin{eqnarray}
&\displaystyle
S_v(u) =0 \,, \quad
S_u(u) = - \frac{C_2}{C_v} \ {\cal S} \left(\frac{e^{-\beta}}{L} \right) \sin{\Psi(u)} \,,
&\nonumber\\
&\displaystyle
S^2(u) = {\cal S} \left(\frac{e^{-\beta}}{L} \right) \ \sin{\Psi(u)}
\,,
&\label{8GW82}\\
&\displaystyle
S^3(u) = {\cal S} \left(\frac{e^{\beta}}{L} \right) \ \cos{\Psi(u)}
 \,.
&\nonumber
\end{eqnarray}
The first and second formulas in (\ref{8GW82}) yield
\begin{equation}
 S^1(u)= \frac{1}{\sqrt2}(S_u{-}S_v) = {-}
\frac{C_2}{\sqrt2 C_v}{\cal S} \left(\frac{e^{{-}\beta}}{L} \right) \sin{\Psi(u)}
 \,. \label{GW821}
\end{equation}
One can see, that due to the spin-axion coupling the longitudinal and the
second transversal components of the spin appear $S^1(u>0)\neq 0$,
$S^2(u>0) \neq 0$. The axionically induced spin rotation is
characterized by the frequency $\Omega(u)=\omega_{25}
\phi^{\prime}(u)$.

\section{Discussion}\label{sec - discussion}

We have established the model of the pseudoscalar (axion)
field action on the spinning particle. In its minimal (curvature independent)
version this model includes the dynamic equation (\ref{211}) and
the equation of the spin evolution (\ref{51}). The terms, which
include the Maxwell tensor $F_{mn}$ and its dual $F^{*}_{ik}$ are
constructed phenomenologically by analogy with (and as
generalization of) the well-known Bargmann-Michel-Telegdi (BMT)
model. As in the BMT model, the dynamic equation (\ref{211}) does
not contain the spin four-vector, and the equation (\ref{51}) is
linear in $S^k$. Is it possible to extend (\ref{51}) by
introducing the spin four-vector quadratically, e.g., as it was made by
Bander and Yee in \cite{B11}? For sure, the scheme, based on the
representations (\ref{1}) of the master equations and on the
decomposition of the basic tensor $\omega^{ik}$ introduced in
(\ref{BMT1}), gives us such tool. For instance, when $\omega^{ik}$ is
linear in the spin, the corresponding dynamic equation is also
linear, and the equation of the spin evolution is quadratic in the
spin four-vector.

However, even in the simplest BMT-like form (\ref{211}) the dynamic equation
includes two new force terms, which points to the axionic extension of the theory.
The first novelty is the term
in which the tensor $eF_{mn}$ is replaced with the tensor $e\phi
F^{*}_{mn}$, where the pseudoscalar multiplier $\phi$ compensates
the pseudo-tensorial nature of the dual Maxwell tensor
$F^{*}_{mn}$. The second novelty is the term with the
gradient-type multiplier $p^l \nabla_l \phi$ in front of
$F^{*}_{mn}$. If the axionic environment indeed produces forces of these kinds,  they
could be tested in high-energy experiments with polarized beams,
e.g. in the LHC. We hope to present the corresponding work and discuss exact
solutions to the whole system of equations of axion
electrodynamics, particle dynamics and spin evolution in the next
paper.

As for this paper, we consider the contribution
from the only new term, which is free of the Maxwell tensor and
linear in the gradient four-vector $\nabla_l \phi$. This term
describes the direct action of the axion field on the particle
spin. This term, $\frac{\omega_{25}}{mc} \nabla_l \phi
\epsilon^{ikln} S_k  p_n$, was introduced phenomenologically, and
the dimensionless coupling constant $\omega_{25}$ should be recognized.
One of the ways to find
$\omega_{25}$ is to make the reduction from the axionically
extended Dirac theory; we will return to this
problem in the future. The second way is to use the analogy with
the axion-photon coupling, which has been considered in Section~\ref{sec - photon polarization}.
If to follow the hypothesis of universality and to compare
the evolutionary equations for the photon
polarization (\ref{g10})  and for the spin rotation (\ref{g107}),
one can assume that $\omega_{25}=\alpha=1$.
Nevertheless, one should repeat that this coupling
constant has to be found experimentally.

One can mention that the term $\frac{\omega_{25}}{m} \nabla_l \phi
\ \epsilon^{ikln} S_k  p_n$ describing the direct spin-axion coupling, can be represented in the BMT-like form.
Indeed, let us take the main term $\frac{e}{m}F^{ik}S_k$ appeared in the BMT model
for the case $g=2$, and consider the decomposition of the Maxwell tensor in the reference frame associated
with the particle moving with the velocity $\frac{p^i}{m}$. In addition, let us assume that in this frame
the electric field $E^k$ is absent, then we obtain that
$F^{ik}=-\frac{1}{m}\epsilon^{ikln}B_l p_n$, where $B^k$ is the four-vector of the corresponding magnetic excitation.
Comparing the terms $\frac{\omega_{25}}{m} \nabla_l \phi
\epsilon^{ikln} S_k  p_n$ and $-\frac{e}{m^2}\epsilon^{ikln}S_k B_l p_n$, we can see that they formally coincide, when
$\omega_{25} \nabla_l \phi = - \frac{e}{m}B_l$. In other words, this analogy hints that the gradient of the pseudoscalar
(axion) field can produce the spin rotation similar to the well-known effect induced by the magnetic field.

Of course, the mentioned analogy is incomplete, since $B^k$ is the space-like four-(pseudo)vector, while the
gradient four-(pseudo)vector $\nabla_k \phi$ can be time-like ($\nabla_i \phi \nabla^i
\phi>0$), space-like ($\nabla_i \phi \nabla^i \phi<0$) or null
($\nabla_i \phi \nabla^i \phi=0$). However, in all three cases, as it was demonstrated using the obtained exact solutions
to the master equations, we deal with the same phenomenon, the axionically induced spin precession.

The typical example for the time-like gradient four-vector $\nabla_k \phi$
is given by the spatially homogeneous cosmological model according to which the pseudoscalar field
corresponds to the relic dark matter axions.
In this case the gradient four-vector reduces to $\delta_i^0 \dot{\phi}$ and the spin rotates in
the plane orthogonal to the direction of the particle motion. The
corresponding time-dependent frequency $\Omega(t){=} \omega_{25}
\dot{\phi}(t) \frac{V}{c}$ ($V$ is the modulus of the
velocity three-vector) is a direct analog of the Larmor frequency.

The static spherically symmetric model of gravitational and axion
fields gives the typical example for the space-like gradient, $\nabla_i \phi
\nabla^i \phi<0$. When the particle moves in radial direction, we
deal again with the spin rotation in the transverse plane. Now it is more
reasonable to speak about the spin turn with respect to radial
variable $r$ rather than with respect to time (see
(\ref{as10})-(\ref{as12})). The rate of spin turn depends on the distance to the center;
in particular, when $p_r(\infty)=0$ and thus $K{=}m{=}E$, we obtain
$\Omega(r){=} \omega_{25}\frac{\phi^{\prime}(r)}{\sqrt{1{-}B(r)}}$. Far from the center
the metric function $B(r)$ has the standard behavior,
$B(r) \to \left(1{-} \frac{2GM}{r}\right)$, so the asymptotic behavior of the frequency $\Omega(r)$,
$\Omega(r\to \infty) \to \frac{\omega_{25}}{\sqrt{2GM}} \sqrt{r} \phi^{\prime}(r)$ is predetermined
by the function $\sqrt{r} \phi^{\prime}(r)$.
One can expect that the zone of strong gravitation gives the maximal contribution into the total turn
of the spin, however, this question should be analyzed in its own right.

When the particle moves along circular orbit around the
axionically active object, one can split the total effect in the
spin rotation into geodesic precession and axionic precession. The
frequency of rotation is given by hybrid formula (\ref{ss76}), and
the axionic frequency $\Omega_{({\rm axion})} {=}
\frac{\omega_{25} {\phi^{\prime}}(R)}{ \sqrt{A(R)}}$ is constant
on the orbit, but depends on the radius of the orbit $R$. The
behavior of the function $\Omega_{({\rm axion})}(R)$ can be
studied only when the distribution of axion field $\phi(r)$ is
found. We hope to return to this question in the next paper in the
context of discussion of qualitative and numerical study of the
total system of master equations.

The case $\nabla_i \phi \nabla^i \phi{=}0$ is typical for the model
with a plane-wave symmetry, for which the axion field depends on the retarded time only,
$\phi(u)$. Again the exact
solutions to the master equations demonstrate the spin rotation
with the frequency $\Omega(u)= \omega_{25} \phi^{\prime}(u)$. Two
cases have to be distinguished in this model. First, when the
particle  moves orthogonally to the front of the gravitational
wave (the so-called longitudinal motion), we deal with simple spin
rotation in the front plane ($S^2 \neq 0$ and $S^3 \neq 0$). When
the projection of the particle momentum on the front plane is
non-vanishing ($p_2\neq 0$, transversal motion), the spin rotation
becomes more sophisticated (the additional, third component of the
spin four-vector appears).

The first obvious conclusion for
all three  applications, is that the
pseudoscalar (axion) field makes the space-time chiral, so that the
left-hand and right-hand rotations of the particle spin
four-vector become non-equivalent. The spin precession
can be indicated as the first general property of the model.

The second general property is that the gravitational field, providing the non-vanishing
gradient of the axion field ($\dot{\phi}\neq 0$, $\phi^{\prime}(r)\neq 0$, $\phi^{\prime}(u) \neq 0$),
activates the spin-axion coupling. In this sense, when the gravity field is strong,
it displays the phenomenon of spin rotation more effectively.

Since the phase of the spin turn is described by the integral formulas of the type
$\Psi = \int d\xi \Omega(\xi)$ ($\xi {=} t$, $\xi {=}r$ or $\xi {=} u$),
the effect of spin rotation is cumulative in the space-time domains, in which the quantities
$\dot{\phi}(t)$, $\phi^{\prime}(r)$ or $\phi^{\prime}(u)$, respectively, hold the sign. In this sense the phase accumulation can be treated as
the third general property of the model.

Finally, we would like to say a couple of words about estimation 
of the described effect. We prefer to do it on the example of relic dark matter axions, which seem to be distributed everywhere, using the model of relativistic charged spin particle motion in a storage ring with magnetic field (see Section~\ref{sec - StorageRing}). In the cosmological context the quantity $\dot{\phi}(t_0)$ can be estimated as $\dot{\phi}(t_0){=}
\frac{1}{\Psi_0} \sqrt{\rho_{({\rm a})}(t_0)}$ using the mass
density of dark matter axions $\rho_{({\rm a})}(t_0)$. Thus, for the
relic cold dark matter axions with the mass density of the order
$\rho_{({\rm DM})} \simeq 0.033 \ M_{({\rm Sun})} {\rm pc}^{-3}$,
for the ultrarelativistic particle with $V \to c$, for the
coupling constant $\frac{1}{\Psi_0}{=} \rho_{{\rm A} \gamma
\gamma} \simeq 10^{-9} {\rm GeV}^{-1}$, we obtain that an optimistic
estimation for the spin rotation frequency (in ${\rm Hz}$) is
$$
\Omega_{({\rm H})} = {\dot{\Psi}}_{({\rm H})} \to \dot{\Psi} \simeq
$$
$$
\simeq 10^{{-}6} \left(\frac{\omega_{25}}{1}\right) \left(\frac{\rho_{{\rm A} \gamma
\gamma}}{10^{{-}9} {\rm GeV}^{{-}1}}\right)\left(\frac{\sqrt{\rho_{({\rm DM})}}}{\sqrt{1.25 {\rm GeV}\cdot {\rm cm}^{{-}3}}}\right) \,.
$$
In order to estimate the possible total axionically induced phase variation, $\Delta \Psi$, for other examples, we have to know the time period, during which the sign of
the frequency is non-changed and is, say, positive. When we deal with homogeneous cosmological model, according to (\ref{Bessel1}),(\ref{Bessel2})
this time period is about of $T=\frac{\pi}{m_{({\rm a})}}$, and thus is negligibly small.
The application to the static spherically symmetric gravitation field is much more promising, since now the derivative $\phi^{\prime}(r)$ is monotonic function, and
the accumulation of the phase of the spin turn can continue during long time for both radial and circular motion of the test particle.

\acknowledgments

AB is grateful to Professor Wei-Tou Ni for fruitful discussion concerning new trends in the physics of axions. This work was supported by Program of Competitive Growth
of KFU (Project 0615/06.15.02302.034), and by Russian Foundation for Basic Research (Grant RFBR N~14-02-00598).

\end{document}